%% file: main.tex
\documentclass[a4paper,11pt]{article}
\usepackage{jinstpub} 
\input{extra.tex}

\title{Quality control of PEN wavelength shifters for DarkSide-20k veto}

\author{Sarthak Choudhary}
\author{for the DarkSide-20k collaboration}
\affiliation{AstroCeNT, Nicolaus Copernicus Astronomical Center of the Polish Academy of Sciences,\\
Rektorska 4, Warsaw, Poland}

\emailAdd{sarthak@camk.edu.pl}

\abstract{Efficient Wavelength Shifters (WLS) are crucial for Liquid Argon (LAr) dark matter detectors. As they grow larger in volume, the scalability of WLS becomes an important concern. Tetraphenyl butadiene (TPB), the most common WLS in use, requires to be deposited with vacuum evaporation impractical for detectors with very large surface area due to its high cost and energy requirements.
The neutron veto of the DarkSide-20k detector will utilise nearly $200$~m$^2$ of polyethylene naphthalate (PEN) wavelength shifter, available in form of large format polymeric foils. In order to assess the quality of PEN sheets in the DarkSide-20k production batch, multiple samples will be tested at cryogenic temperatures.
For this purpose, a new Argon Gas Setup (ArGSet) has been recently commissioned. In this setup, we exploit the Argon scintillation light (128~nm) as excitation for measuring the wavelength shifting efficiency of the samples.
In this work, we will present the results of the first measurements performed at cryogenic temperature with this setup.}


\begin{document}
\maketitle
\flushbottom
\section{Introduction} \label{sec:intro}
DarkSide-20k~\cite{darkside20k} is a Liquid Argon (LAr) detector designed for the direct detection of Weakly Interacting Massive Particles (WIMPs), one of the leading candidates for dark matter. It will utilize a dual-phase liquid argon time projection chamber to achieve unprecedented sensitivity levels to WIMP searches. The main detector will be surrounded by the veto detector filled with LAr. 

LAr detectors work by detecting scintillation photons produced by incident radiation interacting within the LAr volume. The LAr scintillation light is emitted in vacuum ultraviolet (VUV), peaking at 128~nm. Since conventional photodetectors have low efficiency in VUV, a typical scheme for increasing detection efficiency involves covering the inner surfaces of the detector with a wavelength shifting material, such as tetraphenyl butadiene (TPB). Coating large surfaces with vacuum evaporation of TPB is a cost and labour-intensive process. Since Polyethylene naphthalate (PEN)~\cite{kuzniak2019polyethylene} is commercially available in form of large-form sheets, it has been chosen as a scalable alternative to TPB, albeit with approximately $50\%$ reduced wavelength shifting efficiency~\cite{boulay_direct_2021, araujo_r_2022}.

The DarkSide-20k veto detector will employ nearly 200~m$^2$ of PEN wavelength shifter. To ensure strict quality control over commercially procured PEN sheets, Wavelength Shifting Efficiency (WLSE) of multiple PEN samples will be measured before they are installed in the veto detector. For this purpose, a custom setup designated as ArGSet~\cite{choudhary2024cryogenic} will be used, recently developed in collaboration with the National Center for Nuclear Research, \'Swierk.

\section{Quality Assessment Strategy}
It has been observed that the WLSE of PEN may vary from one batch to another, even for the same material grade~\cite{Abraham_2021, araujo_r_2022}. Therefore, it is necessary to ensure the consistency of WLSE of PEN sheets before they get installed in the detector. It is planned to test approximately 50 samples, uniformly extracted from the roll.

Similar to previously done surveys~\cite{kuzniak2019polyethylene}, all samples will be tested with the fastest available technique, using a spectrophotometer equipped with an integrating sphere (at 190~nm wavelength) and at room temperature.
A subset of these samples is to be tested with VUV excitation at room temperature and in cryogenic conditions. 

\section{Setup for cryogenic measurements}
For the cryogenic measurements, we use the Argon Gas Setup (ArGSet)~\cite{choudhary2024cryogenic} consisting of a cryostat, an $^{241}$Am alpha source and two Hamamatsu S14160-6050HS SiPMs~\cite{S14160}.

\begin{figure}[H]
    \centering
        \includegraphics[width = 0.50\textwidth]{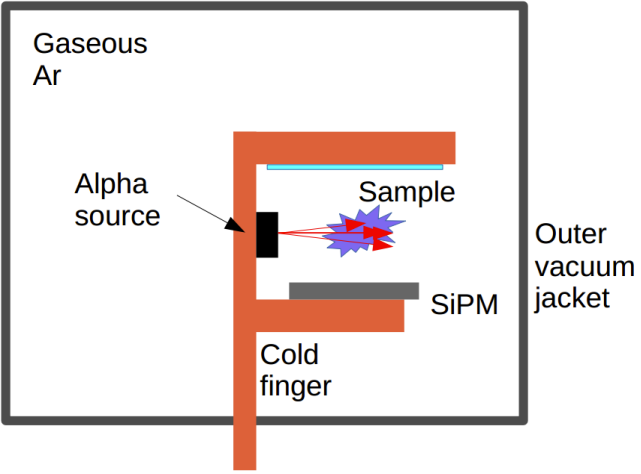}%
        \includegraphics[width = 0.40\textwidth]{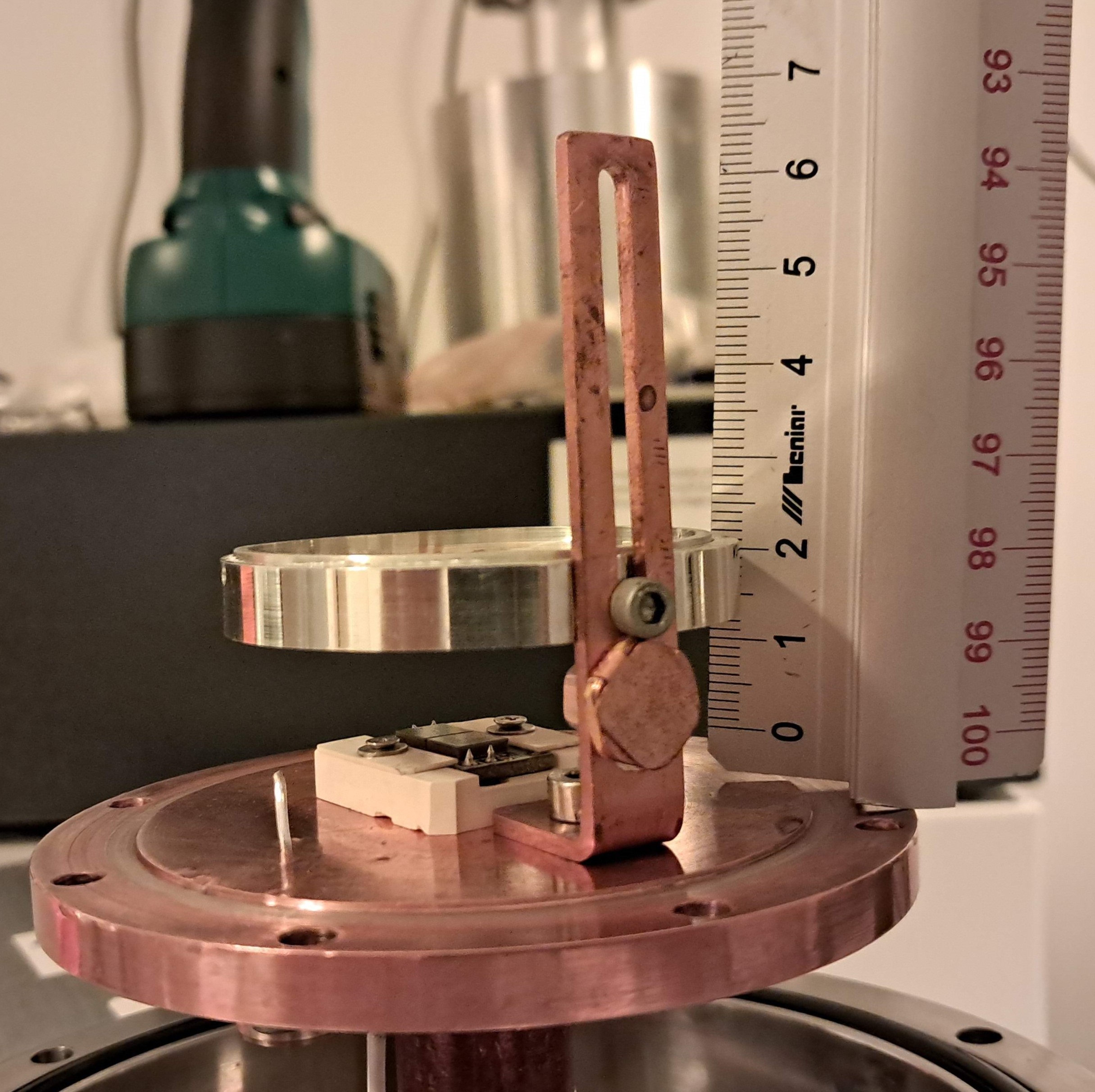}
    \caption{[Left] Schematic of inner chamber of argon gas setup (ArGSet). [Right] Picture of inner chamber; the ruler is placed next to sample holder for comparison.}
    \label{fig:ArGSet}
\end{figure}

In ArGSet, the SiPMs are placed on a thermally conductive boron nitride ceramic base. The ceramic base and the sample holder are connected to a copper cold-finger to ensure the wavelength shifter (WLS) sample and the silicon photomultipliers (SiPMs) remain at cryogenic temperature (figure \ref{fig:ArGSet}).

The inner chamber is vacuum-insulated and is filled with argon gas for the measurements. The alphas from $^{241}$Am induce gaseous argon to scintillate, and these scintillation photons reach the WLS sample. The SiPMs face the sample to detect the fluorescence from WLS (figure~\ref{fig:ArGSet}).

Compared to the previous commissioning tests~\cite{choudhary2024cryogenic}: 1) the distance between the sample and the SiPMs was reduced to maximize the light collection, 2) single photoelectron resolution was achieved following the SiPM gain balancing and calibration, 3) performance and stability of the system over several hours measurement time was studied. 
\section{Setup characterization}
\begin{figure}
\centering
    \includegraphics[width = 0.425\textwidth]{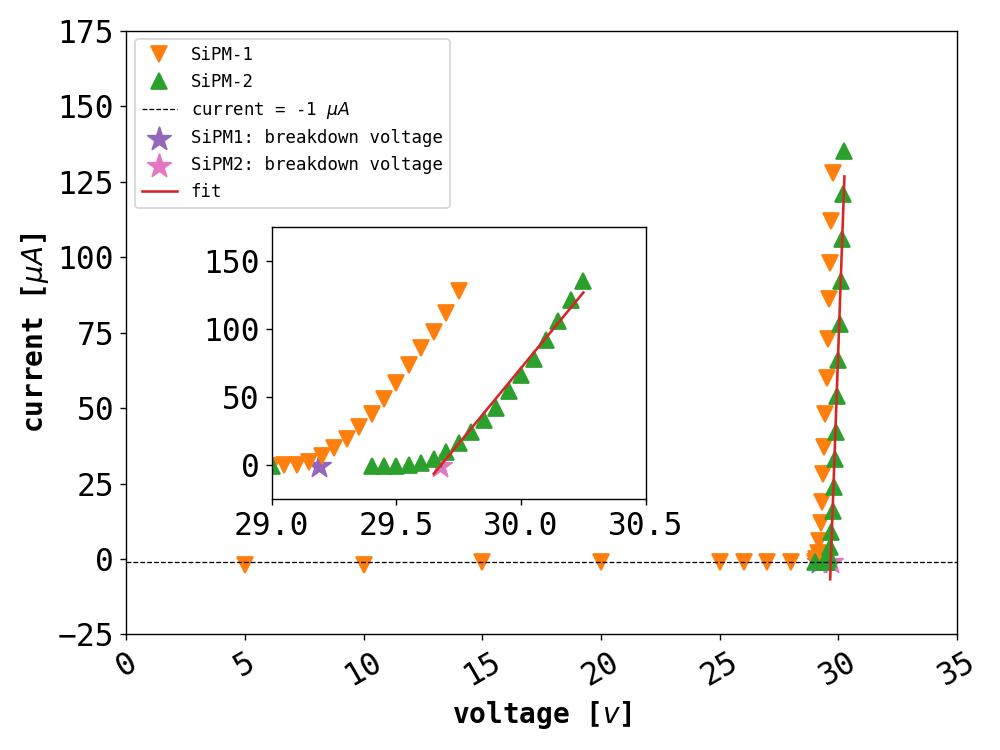}\includegraphics[width = 0.575\textwidth]{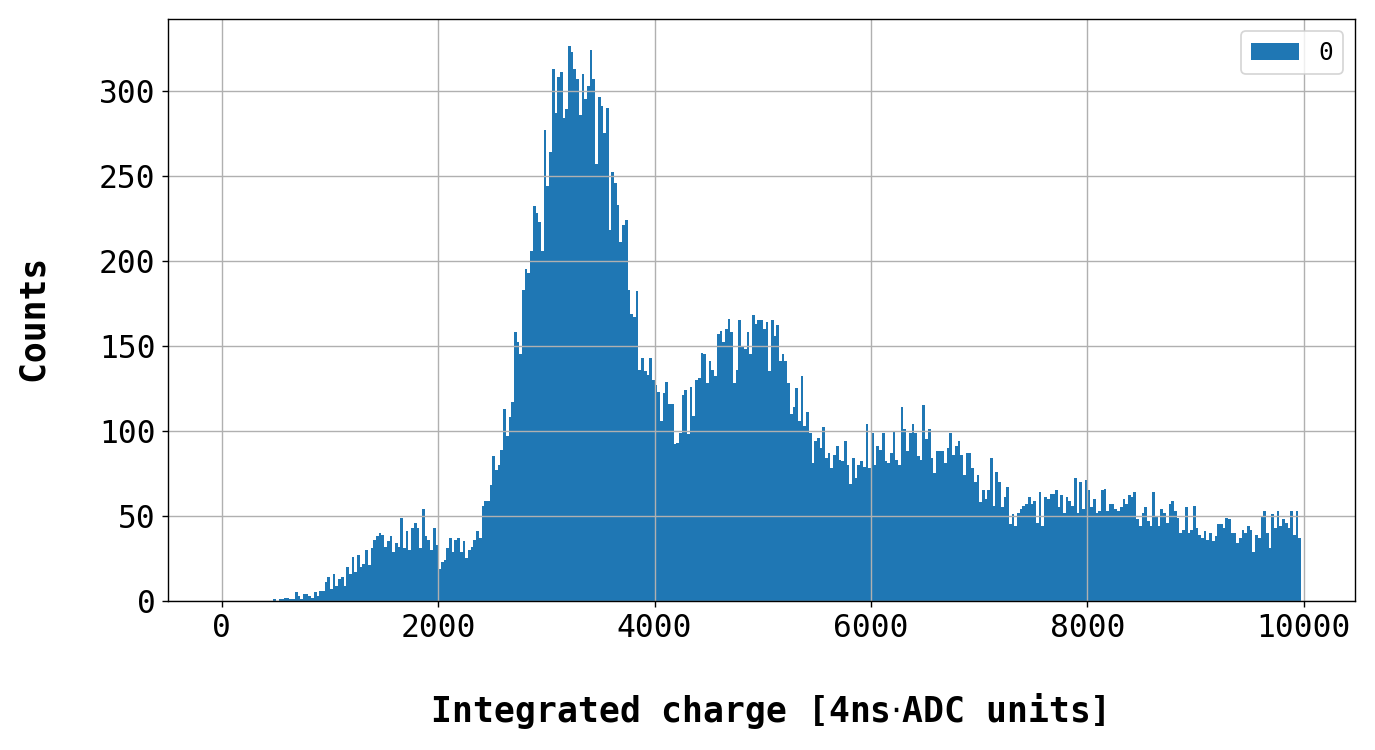}
    \caption{Results of the SiPM characterization. [Left] IV curves for both SiPMs with linear fits extrapolated to breakdown voltages. The inset shows a zoomed view of the plot. [Right] PE charge spectrum for the analogue sum of both SiPM channels (channel 0). The lack of pedestal is caused by only accepting gated single-pulse events above the noise RMS.}
\label{fig:sipm_calibration}
\end{figure}

The setup has individual readout for both SiPM channels (channels 1 and 2) and also provides an analogue sum channel (channel 0) with an additional amplification factor of 5. Throughout this article, channels 0, 1, and 2 are represented by the colors blue, orange, and green, respectively.  
For robustness of the summed output, both SiPMs needs to be operated at the same gain and applying the same overvoltage (OV). 
We find the OV for both SiPMs by measuring the Current-Voltage (IV) response curve, as shown in figure~\ref{fig:sipm_calibration} (left) along with a fit. 
The breakdown voltage is found by fitting a straight line to the steep part of IV curves and extrapolating it to the crossing point with the baseline. Resulting breakdown voltages are $29.19 \pm 0.06$~V and $29.67 \pm 0.06$~V, respectively, for SiPM-1 and SiPM-2.
The SiPM calibration procedure is performed with a hardware dedicated configuration of the setup: in vacuum, in absence of the inner chamber, and with a dedicated outer vacuum shroud supplied with a fibre optics feedthrough. To provide enough illumination for the IV curve measurement, the feedthrough is kept open to ambient light.

Subsequently, the single photoelectron (PE) response calibration is performed, using a triggered LED driver providing 400~nm light pulses with 8~ns duration (CAEN~SP5601), connected via fibre optics.
SiPMs are kept at cryogenic temperature. Pulses gated with the LED trigger were collected, and to further reduce the noise, a matched filter~\cite{tim_darkside} tuned to the characteristics of genuine SiPM pulses was applied to the data. Additionally, events inconsistent with a single pulse above the noise RMS were removed. The integrated PE charge spectrum (fingerplot) made from remaining events is shown in the figure~\ref{fig:sipm_calibration} (right).

\begin{figure}
    \centering
    \includegraphics[width=0.75\linewidth]{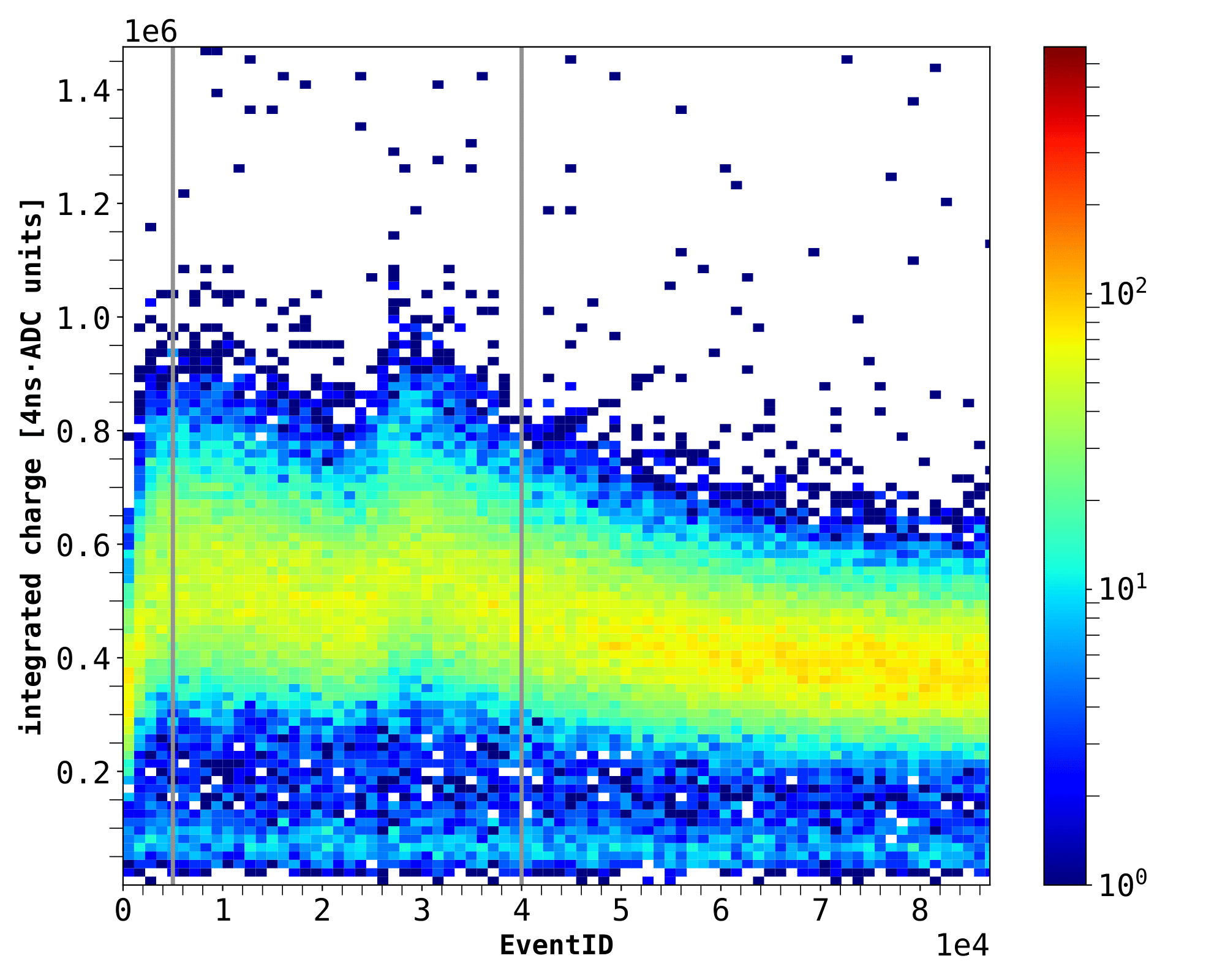}
    \caption{Time evolution of integrated charge (proportional to the amount of detected photons) in the analogue sum channel. Grey vertical lines indicate the event range selected for analysis. This event range is equivalent to approximately 12~hours of measurement. The second peak corresponds to topping up of the cold finger liquid nitrogen reservoir.}
    \label{fig:ly_time_evolution}
\end{figure}

It was observed that the integrated charge (the sum of raw waveforms) drifts over time (see figure \ref{fig:ly_time_evolution}), following the gradual warm-up of the setup. Therefore, we only use a range of collected events (see caption of figure \ref{fig:ly_time_evolution}).

\section{Measurements}
While the basic operation of the setup has already been demonstrated~\cite{choudhary2024cryogenic}, in this work a more in-depth systematic study is reported, motivating further improvements to the setup to fulfil the requirements for the quality assessment campaign.

The PEN sample was loaded in the sample holder, and the cell was flushed with gaseous argon (6N purity), valved off, enclosed in insulating vacuum, and cooled down.

As the measurements suffered from occasional intermittent noise events, simple data quality event selection cuts were introduced. The RTD temperature sensor, identified as one of the main sources of noise, was therefore only used at the beginning and at the end of each run. The temperature for the present measurement was -167.01 $^{\circ}$C in the beginning and -171.20 $^{\circ}$C at the end of the measurement. Note that the full data set was not used for the analysis and the temperature would have been lower for the range selected.

\begin{figure}[htpb]
    \centering
    \includegraphics[width=0.75\linewidth]{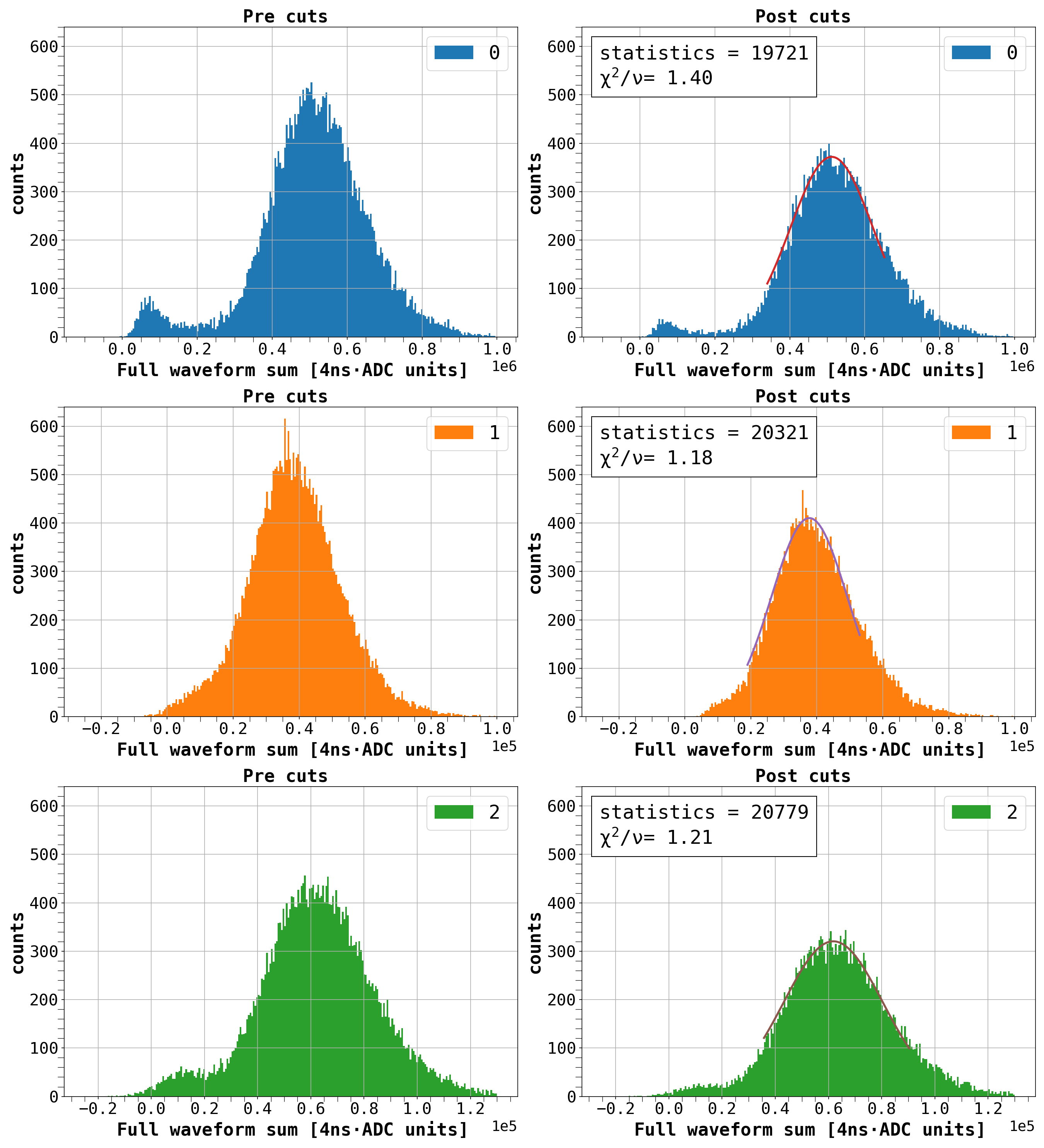}
    \caption{Charge distribution before and after applying event selection cuts. The integrated charge is obtained by summing up the filtered waveforms in 4~ns time bins. The fits to charge distribution are shown in the plots on the right.}\label{fig:charge_distribution}
\end{figure}

\begin{figure}[htpb]
    \centering
    \includegraphics[width=0.75\linewidth]{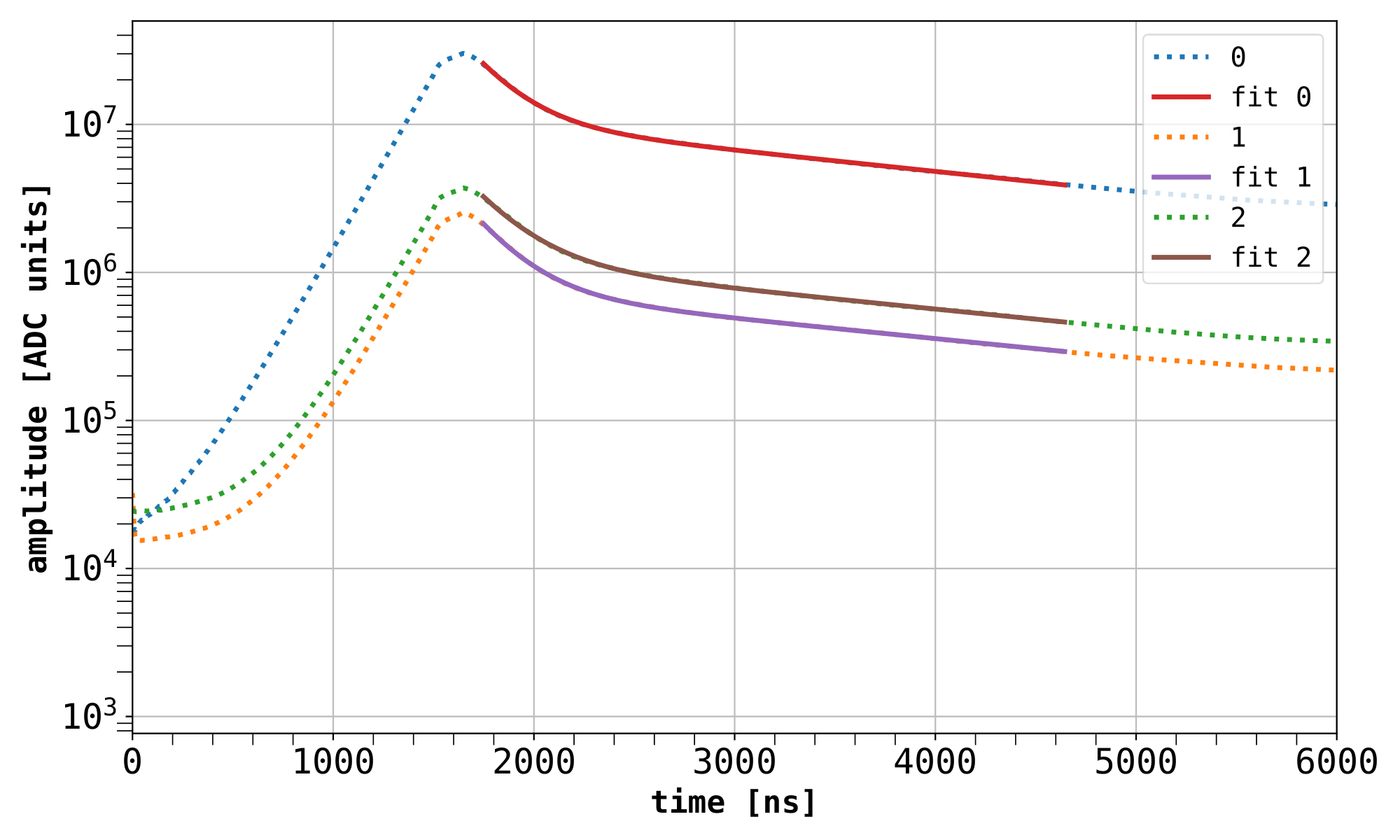}
    \caption{Double exponential fit to the stacked waveform. The mean argon triplet lifetime calculated from fit is $3.06 \pm 0.16$ (syst.) $\mu$s.}
    \label{fig:argon_lifetime}
\end{figure}

Figure~\ref{fig:charge_distribution} shows the comparison of charge distributions before (left) and after (right) applying the event selection cuts for the PEN measurement run. The cuts remove events with a) excessive charge in the pre-trigger window, b) charge weighted average time bins of the waveform inconsistent with the value expected from gaseous argon scintillation, and c) excessive time difference between pulses in both SiPM channels.
A Gaussian fit to the charge distribution for each channel was then performed, as shown in the plots in figure \ref{fig:charge_distribution} (right). We then extracted the mean charge from these fits, corresponding to the $\alpha$-induced scintillation events.
The table~\ref{table:nph} lists the mean value of charge collected, and the number of photoelectrons estimated from that charge value.
\begin{table}[H]
    \centering
    \caption{Number of detected Photoelectrons (PE)}
    \label{table:nph}
    \begin{tabular}{|>{\centering\arraybackslash}m{5.0cm}|>{\centering\arraybackslash}m{5.0cm}|}
        \hline
               & analogue sum channel \\
        \hline
        Mean Integral Charge [4ns$\cdot$ADC units]    & 5.184e+05   \\
        \hline
        sPE charge [4ns$\cdot$ADC units]  & 1612\\
        \hline
        Photoelectrons N$_{pe}$    & 321 \\
        \hline
    \end{tabular}
\end{table}
The figure \ref{fig:argon_lifetime} shows the fit to the stacked waveforms. We use the expression in equation \ref{eq: lifetime_func} for fitting the stacked waveform. The parameters $\tau_3$ and $\tau_1$ give the time constant for the slow and fast component of the scintillation pulse, respectively. 
\begin{equation}
    f_{1}(x) = a_3 e^{-(x-a_0)/\tau_3} + a_1 e^{-(x-a_0)/\tau_1} \label{eq: lifetime_func}
\end{equation}
    
The triplet (slow component) lifetime found from the fit is $3.06 \pm 0.16$ (syst.) $\mu$s, significantly improved with respect to commissioning ~\cite{choudhary2024cryogenic}, and stable over the course of entire run.  The measured value of triplet lifetime is consistent with reported values in literature, varying from 2.8 to 3.2 $\mu$s~\cite{AkashiRonquest2019}. 

\section{Summary and future work}
The strategy for quality control of the PEN wavelength shifter is discussed. A gas argon setup operating at cryogenic temperature is presented for measuring wavelength shifting efficiencies. Though the setup suffers from intermittent noise, it is efficiently handled with data quality event selection cuts. The single PE charge and argon triplet lifetime are measured. The setup is shown to be capable of achieving single photoelectron resolution, and detecting order of a hundred photoelectrons generated from a PEN wavelength shifter sample.

The collected event charge appears to drift during a run in correlation with the temperature. The stability of the triplet lifetime over several hours excludes outgassing as the cause for the observed drift. 
Consequently, we can only utilize a subset of the data collected within a specific time window, typically around 10~hours. Improvements to temperature monitoring, aimed at offline correction of SiPM gains, are planned.

Furthermore, while some asymmetry between the two SiPM signals is expected as the result from the setup’s geometry, this asymmetry has been observed to vary between runs, potentially due to inconsistent argon density levels during different runs, affecting the geometry of the energy deposit and scintillation light emission. To address this, a cryogenic-rated pressure gauge (Kulite CTL-190SM-30A) is being installed to more accurately monitor the gas pressure. Control of the systematic effects is expected to improve significantly after the pressure gauge is installed.
\acknowledgments
This work was funded by the National Science Centre, Poland (Grant No. 2022/47/B/ST2/02015),
by the International Research Agenda Programme AstroCeNT (MAB/2018/7) funded by the Foundation for Polish Science from the European Regional Development Fund, and from the EU’s
Horizon 2020 research and innovation programme under grant agreement No 952480 (DarkWave). Tadeusz Sworobowicz is sincerely thanked for his valuable assistance with the ArGSet.
\bibliographystyle{JHEP}
\bibliography{references}
\end{document}

%% file: extra.tex
\usepackage{caption}
\usepackage{subcaption}
\usepackage{float}
\usepackage{array}
\usepackage{svg}